\documentclass[twocolumn,preprintnumbers,amsmath,amssymb]{revtex4-1}

\usepackage{array}
\usepackage[dvipdfmx]{graphicx}
\usepackage{tabularx}
\usepackage{longtable}
\usepackage{dcolumn} 
\usepackage{bm}
\usepackage{color}

\begin{document}

\title{Analytical expression of negative differential thermal resistance in a macroscopic heterojunction}

\author{Wataru Kobayashi}
\email{kobayashi.wataru.gf@u.tsukuba.ac.jp}
\affiliation
{Division of Physics, Faculty of Pure and Applied Sciences, University of Tsukuba, 
Ibaraki 305-8571, Japan}
\affiliation
{Tsukuba Research Center for Energy Materials Science {\rm (TREMS)}, 
University of Tsukuba, Ibaraki 305-8571, Japan}

\begin{abstract}

Heat flux ($J$) generally increases with temperature difference in a material. 
A differential coefficient of $J$ against temperature ($T$) is called differential thermal conductance ($k$), 
and an inverse of $k$ is differential thermal resistance ($r$). 
Although $k$ and $r$ are generally positive, they can be negative in a macroscopic heterojunction 
with positive $T$-dependent interfacial thermal resistance (ITR). 
The negative differential thermal resistance (NDTR) effect is an important effect that can 
realize thermal transistor, thermal memory, and thermal logic gate.  
In this paper, we examine analytical expressions of $J$, 
$k$, $r$, and other related quantities as a function of parameters related to 
thermal conductivity ($\kappa$) and ITR in a macroscopic heterojunction 
to precisely describe the NDTR effect. 

\end{abstract}

\maketitle
\section{Introduction}

\begin{figure}[b]
\begin{center}
\includegraphics[width=60mm]{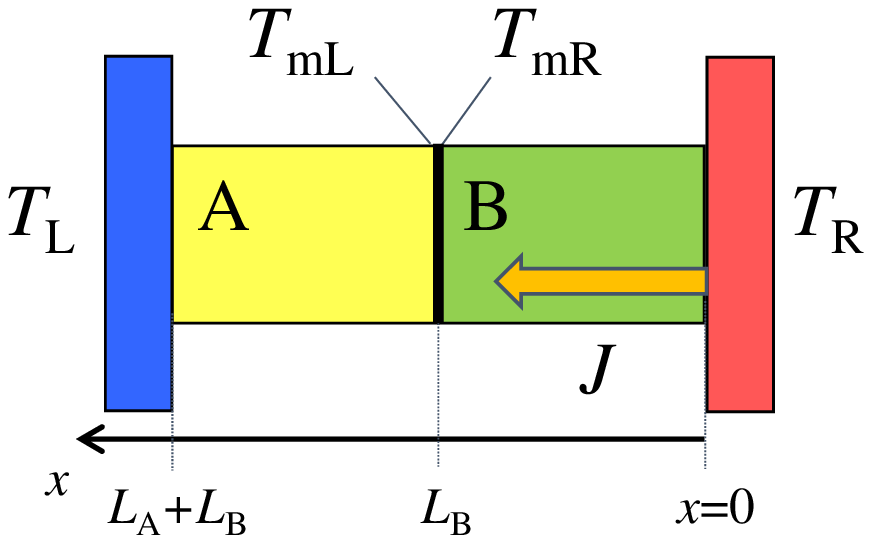}
\caption{(Color online) 
Schematic figure of a macroscopic heterojunction consists of juxtaposing material A and material B 
with interfacial thermal resistance (ITR) 
and non-uniform thermal conductivities ($\kappa$) against position ($x$) 
and temperature ($T$). 
The edge (at $x=0$) of the material B 
is contacted with a heat bath with right-hand side (denoted by R) high temperature ($T_{\rm R}$), 
and the heat flux ($J$) 
flows to a heat bath with left-hand side (L) low temperature ($T_{\rm L}$) 
[$T_{\rm L}<T_{\rm R}$] at $x=L_{\rm A} + L_{\rm B}$ 
where an edge of the material A is contacted. At $x=L_{\rm B}$, the materials A and B 
are connected, in which $T$-dependent ITR is introduced. 
The right(left)-hand side temperature at $x=L_{\rm B}$ 
is denoted by $T_{\rm mR}$($T_{\rm mL}$).
}
\label{f1}
\end{center}
\end{figure}

Thermal control is recently attracted much attention to address 
worldwide challenges such as energy harvesting, carbon neutral, 
warming temperatures, smart society, and sustainable development goals. 
The thermal-control technology consists of heat conduction, energy conversion, 
cooling, thermal storage, heat insulating, and thermal radiation technologies. 
Further, focusing on the heat conduction, 
thermal-circuit elements such as thermal rectifier, 
and thermal transistor, as a counterpart of electronic-circuit elements, 
are important to precisely control heat flux ($J$) \cite{li3,ding1}. 

Thermal rectifier is an analogue of electrical rectifier, in which the heat flux in a forward direction 
is larger than that in the reverse direction. Theoretical calculations 
on the thermal rectification in microscopic one-dimensional system were reported \cite{terraneo1,li1,li4}. 
In agreement with the theories, a thermal rectification in 
a carbon nanotube with mass gradient was demonstrated \cite{chang1}. 
After that, a design of a bulk thermal rectifier was proposed \cite{peyrard1}. 
In fact, the thermal rectification was demonstrated in bulk oxides \cite{kobayashi1}. 
Thus, both microscopic and macroscopic theories have successfully lead experimental 
realizations in both microscopic and macroscopic systems 
\cite{li3,terraneo1,li1,chang1,peyrard1,kobayashi1,yang1,sawaki1}.  

Negative differential thermal resistance (NDTR) is a key effect which realizes 
thermal transistor \cite{li2,lo1}, thermal logic gate \cite{wang1}, 
and thermal memory \cite{wang2}. 
In the thermal transistor, $J$ can be amplified. An amplification factor ($\gamma$) defined by 
\begin{equation}
\gamma \equiv \left|\frac{r_{\rm s}}{r_{\rm s} +r_{\rm d}}\right|, 
\label{1}
\end{equation} 
where $r_{\rm s}$ and $r_{\rm d}$ represent differential thermal resistances ($r$) at source and drain, respectively, 
becomes to be above one when $r_{\rm s}$ or $r_{\rm d}$ is negative \cite{li3,li2}. 
Thus, many theoretical efforts have been done to realize the NDTR effect. 
First, Li {\it et al.} investigated one-dimensional Frenkel-Kontorova (FK) lattice model and 
found NDTR effect\cite{li2}. Then, one dimensional atomic lattice models with 
mass gradient, two segment, different interactions, and/or on-site potentials were 
widely investigated and the NDTR effects were found by these theories 
\cite{yang3,lo1,hu1,chen1,he1,shao1}. 
He {\it et al.} found that the origin of NDTR consists in the competition between 
temperature difference and a negative temperature dependence of thermal 
boundary conductance in a chain of two weakly coupled nonlinear lattices \cite{he1}. 
Shao {\it et al.} found that the NDTR effect highly depends on the properties of the 
interface and the system size in the two segment FK model \cite{shao1}. 
Although NDTR effects were also found in graphene nanoribons and heterojunction nanoribons, 
as the length of the nanoribons increases, unfortunately the NDTR effects gradually disappear 
\cite{hu1,chen1}.   
Thus, an experimental realization of the NDTR effect 
seems difficult to treat nano-scale objects with proper interfacial properties. 
Indeed, the NDTR effect has not been experimentally observed yet. 

A bulk NDTR effect is promising for applicational points of view. 
Recently, Yang {\it et al.} theoretically found the bulk NDTR effect in a macroscopic 
homojunction with interface \cite{yang2}. 
The NDTR element consists of juxtaposing bulk materials (materials A and B) with 
interface with interfacial thermal resistance (ITR) as shown in Fig. \ref{f1}. 
When ITR exhibits a certain temperature dependence, the macroscopic homojunction present 
the bulk NDTR effect. Although they revealed the specific temperature dependence of ITR is 
essential to exhibit the bulk NDTR effect, they did not show precise analysis of this phenomenon. 

In this paper, we investigate analytical expressions of the NDTR effect to 
understand the NDTR effect more precisely. 
$J$, $k$, $r$, and other related quantities are analytically described as a function 
of several parameters related to thermal conductivity ($\kappa$) and ITR in a macroscopic heterojunction. 

\begin{figure}[t]
\begin{center}
\includegraphics[width=70mm]{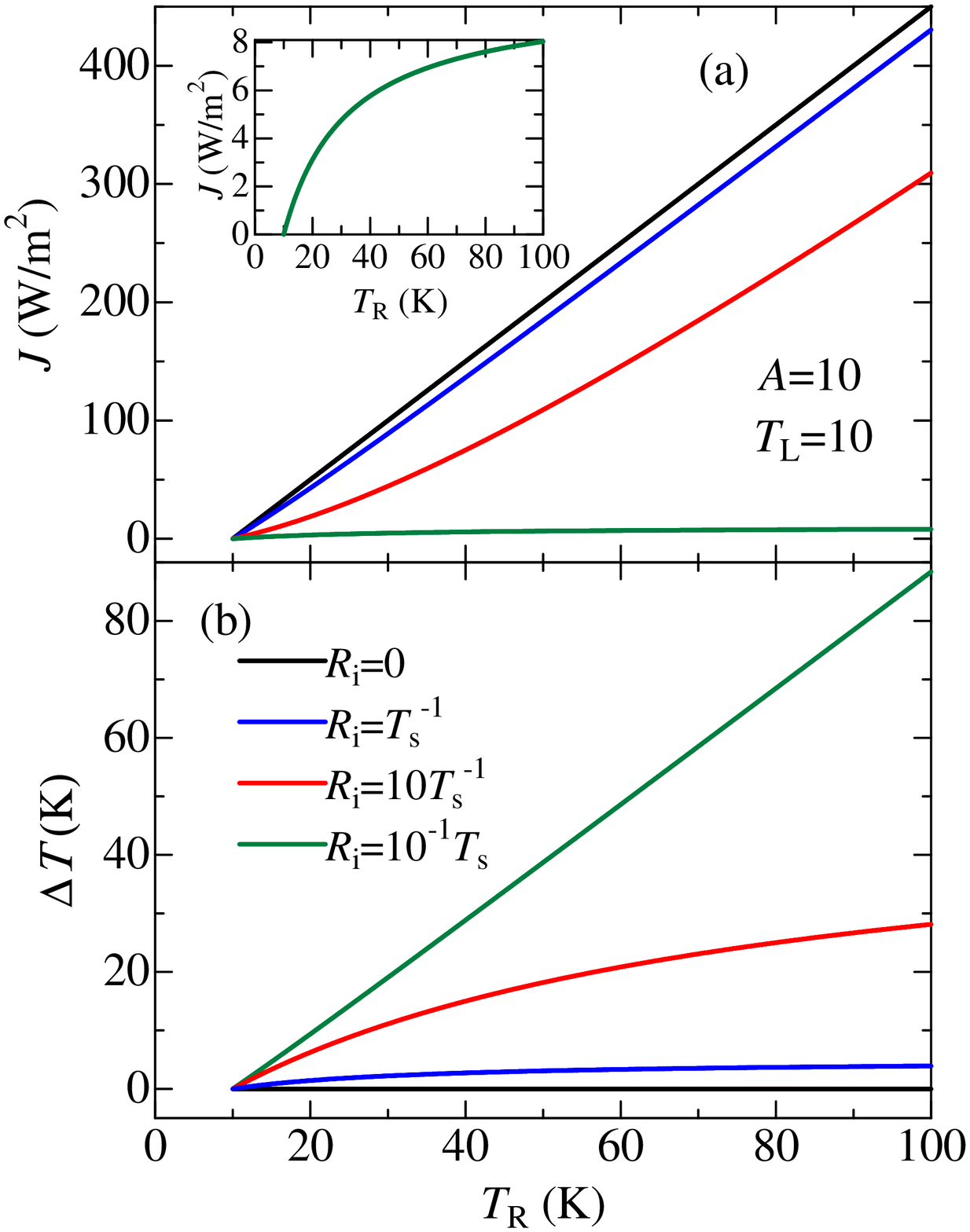}
\caption{(Color online) 
(a) Interfacial thermal resistance ($R_{\rm i}$) dependence of $J$ against 
right-hand side temperature ($T_{\rm R}$) at high-$T$ heat bath. 
A sum of $T_{\rm mL}$ and $T_{\rm mR}$ is defined as 
$T{\rm _s}$ [$T_{\rm s}\equiv T_{\rm mL}+T_{\rm mR}$], where $T_{\rm mL}$ ($T_{\rm mR}$) represents 
a left(right)-hand side temperature at the interface.
(b) $R_{\rm i}$ dependence of temperature difference ($\Delta T \equiv T_{\rm mR}-T_{\rm mL}$) 
in between $T_{\rm mL}$ and 
$T_{\rm mR}$ at the interface against $T_{\rm R}$. 
Lines in Figs. \ref{f2} (a) and (b) represent analytical expressions from Eqs. 
\ref{11} and \ref{12}, respectively. 
Parameters $A$ and $T_{\rm L}$ are fixed to be 10 and 10, respectively. 
}
\label{f2}
\end{center}
\end{figure} 

\begin{figure}[t]
\begin{center}
\includegraphics[width=70mm,clip]{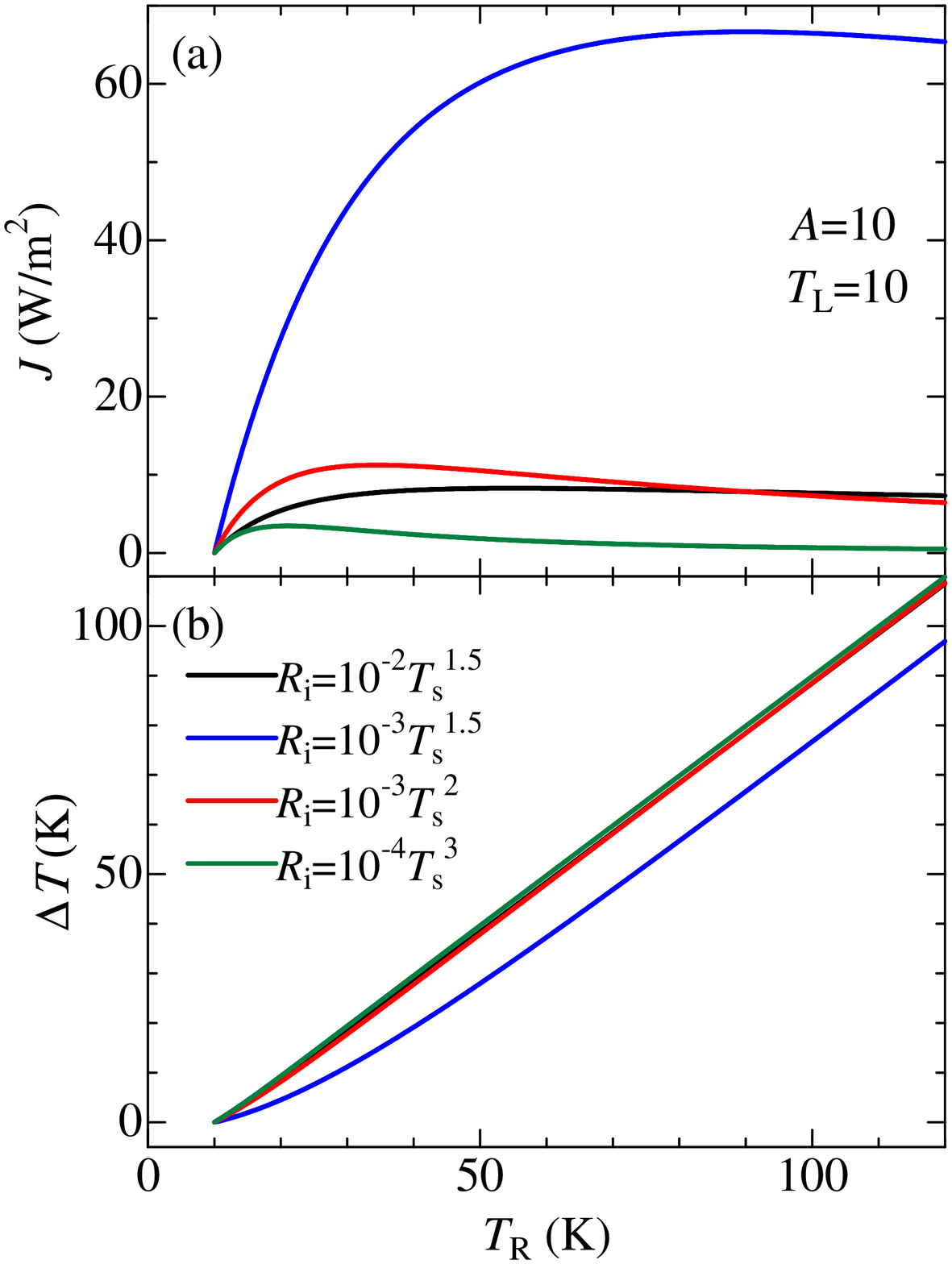}
\caption{(Color online) 
(a) $R_{\rm i}$ dependence of $J$ against $T_{\rm R}$. 
(b) $R_{\rm i}$ dependence of $\Delta T$ against $T_{\rm R}$. 
Lines in Figs. \ref{f3} (a) and (b) represent analytical expressions from Eqs. 
\ref{11} and \ref{12}, respectively. 
Parameters $A$ and $T_{\rm L}$ are fixed to be 10 and 10, respectively. 
}
\label{f3}
\end{center}
\end{figure} 

\begin{figure*}[t]
\begin{center}
\includegraphics[width=130mm,clip]{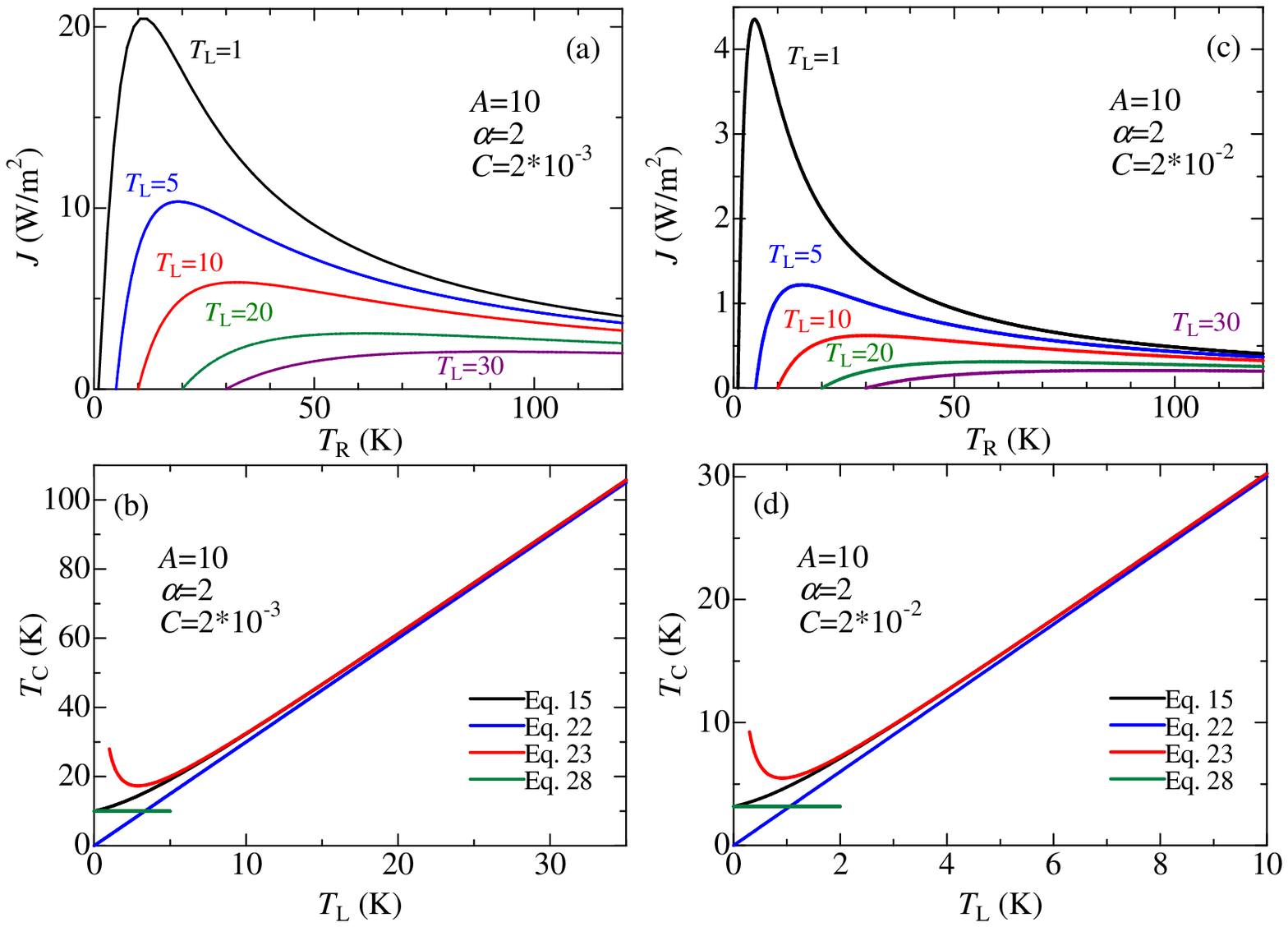}
\caption{(Color online) 
Parameter $T_{\rm L}$ dependence of $J$ against $T_{\rm R}$ at (a) $A=10$, $\alpha=2$, and 
$C=2\times 10^{-3}$, (c) $A=10$, $\alpha=2$, and $C=2\times 10^{-2}$. 
A temperature ($T_{\rm c}$) at $k \equiv \frac{\partial J}{\partial T_{\rm R}}=0$ against $T_{\rm L}$ 
described by analytical expressions Eqs. \ref{15}, \ref{22}, \ref{23}, and \ref{28} at (b) $A=10$, 
$\alpha=2$, and $C=2\times 10^{-3}$, and at (d) $A=10$, $\alpha=2$, and $C=2\times 10^{-2}$. 
}
\label{f4}
\end{center}
\end{figure*} 

\begin{figure}[t]
\begin{center}
\includegraphics[width=70mm,clip]{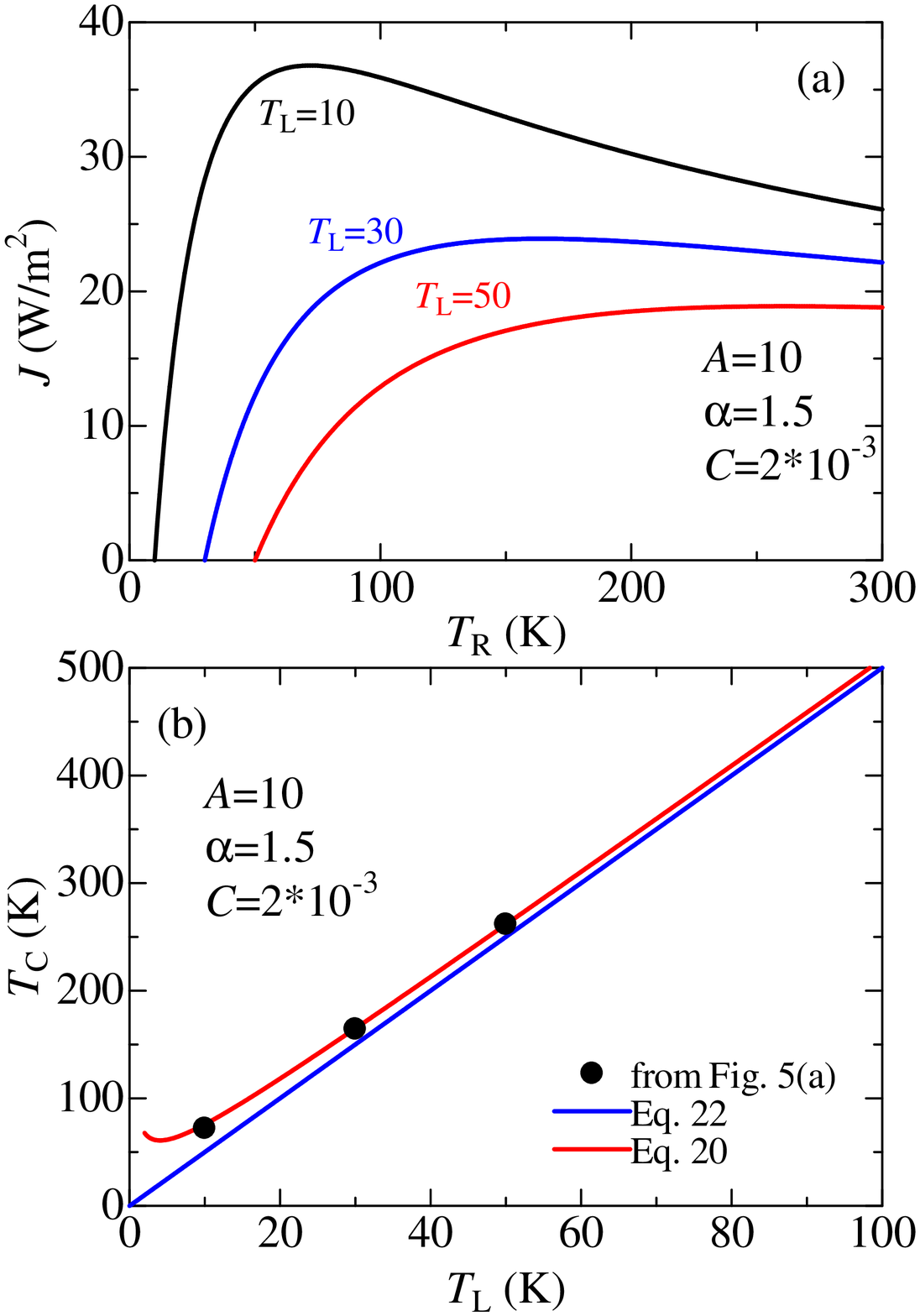}
\caption{(Color online) 
(a) $T_{\rm L}$ dependence of $J$ against $T_{\rm R}$ at $A=10$, $\alpha=1.5$, and 
$C=2\times 10^{-3}$, and (b) $T_{\rm c}$ against $T_{\rm L}$ 
described by analytical expressions Eqs. \ref{20}, 
and \ref{22} at $A=10$, $\alpha=1.5$, and $C=2\times 10^{-3}$. Dots are plotted 
from temperatures at $k=0$ of the data in Fig. \ref{f5}(a).
}
\label{f5}
\end{center}
\end{figure}

\section{methods}

Fourier's law is a fundamental law for describing macroscopic heat conduction in condensed matter, 
which is derived from phenomenological equations \cite{onsager1}.
In this paper, 
we assume insulated one-dimensional system consists of juxtaposing material A with the length 
of $L_{\rm A}$ and material B with the length of $L_{\rm B}$ with interface as shown in Fig. \ref{f1}. 
The interface has temperature dependent interfacial thermal resistance ($R_{\rm i}$). 
At the interface, temperature difference occurs due to $R_{\rm i}$. 
A left(right)-hand side temperature at the interface is $T_{\rm mL}$ ($T_{\rm mR}$) 
[m denotes middle].
Both the materials exhibit non-uniform thermal conductivity 
against $x$ and $T(x)$.
We use Fourier's law written as 
\begin{equation}
J=-\kappa[x, T(x)]\frac{dT(x)}{dx}.
\label{2}
\end{equation}
Since time derivative of internal energy density ($u$) is zero at steady state, 
$\nabla \cdot J =0$ is obtained from the energy conservation law. 
Note that radiation loss is ignored in this paper. 
Thus, $J$ becomes constant at any position in the one-dimensional system. 

Then, integral of $J$ with respect to $x$ in the material B is shown below, 
\begin{equation}
\int_0^{L_{\rm B}}J dx = \int_0^{L_{\rm B}}-\kappa_{\rm B}[x, T(x)]\frac{dT(x)}{dx}dx = 
\int_{T_{\rm mR}}^{T_{\rm R}} 
\kappa_{\rm B}dT,
\label{3}
\end{equation}
where $\kappa_{\rm B}$, $T_{\rm R}$, and $T_{\rm mR}$ represent, 
$\kappa$ of the material B, a temperature at right-hand side high-$T$ heat bath, and 
a right-hand side temperature at the interface ($x=L_{\rm B}$), respectively. 
Similarly, the integral in the material A is shown below, 
\begin{equation}
\begin{split}
\int_{L_{\rm B}}^{L_{\rm A}+L_{\rm B}}Jdx &= \int_{L_{\rm B}}^{L_{\rm A}+L_{\rm B}}
-\kappa_{\rm A}[x, T(x)]\frac{dT(x)}{dx}dx\\ 
&= \int_{T_{\rm L}}^{T_{\rm mL}}\kappa_{\rm A}dT,
\end{split}
\label{4}
\end{equation}
where $\kappa_{\rm A}$, $T_{\rm L}$, and $T_{\rm mL}$ are $\kappa$ of the material A, 
a temperature at left-hand side low-$T$ heat bath, and a left-hand-side temperature 
at the interface, respectively. 

Then, we introduce an interface with $R_{\rm i}$. 
$J$ at the interface is describes as 
\begin{equation}
J =\frac{T_{\rm mR}-T_{\rm mL}}{R_{\rm i}(T_{\rm mL}, T_{\rm mR})}.
\label{5}
\end{equation}

Since $J$ is constant at any position of $x$, Eq. \ref{5} is equal to Eqs. \ref{3} and \ref{4}. 
Thus, 
\begin{equation}
J = \frac{1}{L_{\rm A}}\int_{T_{\rm L}}^{T_{\rm mL}} \kappa_{\rm A}dT=\frac{T_{\rm mR}-T_{\rm mL}}{R_{\rm i}
(T_{\rm mL}, T_{\rm mR})} = \frac{1}{L_{\rm B}}\int_{T_{\rm mR}}^{T_{\rm R}} \kappa_{\rm B}dT
\label{6}
\end{equation}
is obtained. 

In this paper, as Yang {\it et al.} used \cite{yang2}, we assume power law as temperature dependence of 
ITR,  
\begin{equation}
R_{\rm i}(T_{\rm mL}, T_{\rm mR})=C(T_{\rm mR}+T_{\rm mL})^{\alpha}, 
\label{7}
\end{equation}
where $C$ is constant which regulates the magnitude of ITR, $\alpha$ is constant which 
regulates the power, and the sum $T_{\rm s} \equiv T_{\rm mR}+T_{\rm mL}$ means mean temperature. 
As shown by Yang {\it et al.}, when $\alpha>1$, the NDTR effect is occurred. This condition is 
easily derived by searching a condition that the $T_{\rm mR}$ derivative of $J$ is zero 
($\frac{\partial J}{\partial T_{\rm mR}}=0$) shown below, 
\begin{equation}
T_{\rm mR}=\frac{\alpha+1}{\alpha-1}T_{\rm mL}.
\label{8}
\end{equation}
$\alpha>1$ is essential for positively reasonable solution $T_{\rm mR}>T_{\rm mL}>0$.

To solve Eq. \ref{6}, here, both $\kappa_{\rm A}$ and $\kappa_{\rm B}$ are set to be constants 
as zeroth-order approximation. 
In addition, we set $\frac{\kappa_{\rm A}}{L_{\rm A}}=\frac{\kappa_{\rm B}}{L_{\rm B}}=A$. 
Then, two equations are derived from Eq. \ref{6} to obtain $T_{\rm mL}$ and $T_{\rm mR}$ as follows, 
\begin{equation}
\begin{split}
&A(T_{\rm mL}-T_{\rm L})= A(T_{\rm R}-T_{\rm mR}), \\
&\frac{T_{\rm mR}-T_{\rm mL}}{C(T_{\rm mR}+T_{\rm mL})
^{\alpha}}=A(T_{\rm R}-T_{\rm mR}).
\end{split}
\label{9}
\end{equation}

These polynomial equations can be analytically solved, and 
the both solutions of $T_{\rm mR}$ and $T_{\rm mL}$ are obtained. 
Then all the quantities $T_{\rm mR}$, $T_{\rm mL}$ ($=T_{\rm R}+T_{\rm L}-T_{\rm mR}$), 
$J$, $\Delta T$, and $k$ ($=r^{-1}$) 
are easily derived as a function of $A$, $C$, $\alpha$, $T_{\rm L}$, and $T_{\rm R}$ shown below, 

\begin{equation}
T_{\rm mR}=\frac{AC(T_{\rm R}+T_{\rm L})^{\alpha}T_{\rm R}+(T_{\rm R}+T_{\rm L})}{2+AC(T_{\rm R}+T_{\rm L})^{\alpha}},
\label{10}
\end{equation}

\begin{equation}
J=A\left(T_{\rm R}-T_{\rm mR}\right)=\frac{A(T_{\rm R}-T_{\rm L})}{2+AC(T_{\rm R}+T_{\rm L})^{\alpha}},
\label{11}
\end{equation}

\begin{equation}
\Delta T\equiv T_{\rm mR}-T_{\rm mL}=\frac{AC(T_{\rm R}+T_{\rm L})^{\alpha}(T_{\rm R}-T_{\rm L})}{2+AC(T_{\rm R}+T_{\rm L})^{\alpha}},
\label{12}
\end{equation}

\begin{equation}
\begin{split}
k (=r^{-1})&\equiv \frac{\partial J}{\partial T_{\rm R}}\\
&=\frac{A}{2+AC(T_{\rm R}+T_{\rm L})^{\alpha}}\left(1-\frac{\alpha \left(\frac{T_{\rm R}-T_{\rm L}}{T_{\rm R}+T_{\rm L}}\right)}{1+\frac{2}{AC(T_{\rm R}+T_{\rm L})^{\alpha}}}\right).
\label{13}
\end{split}
\end{equation}
Considering $J \geq 0$ and $\Delta T \geq 0$, $T_{\rm R} \geq T_{\rm L}$ is naturally derived. 
At $k=0$, an equation is derived from Eq. \ref{13} shown below, 
\begin{equation}
AC(T_{\rm R}+T_{\rm L})^{\alpha-1}\{(\alpha-1)T_{\rm R}-(\alpha+1)T_{\rm L}\}=2>0. 
\label{14}
\end{equation}
Thus, a condition $(\alpha-1)T_{\rm R}-(\alpha+1)T_{\rm L}>0$ must be realized, which 
leads $T_{\rm R}>\frac{\alpha+1}{\alpha-1}T_{\rm L}$ to observe NDTR. 
Eq. \ref{14} can be analytically solved at $\alpha=2$, and the solution is obtained as 
\begin{equation}
T_{\rm c} \equiv T_{\rm R} (k=0) = T_{\rm L}+\sqrt{4T_{\rm L}^2+\frac{2}{AC}}>3T_{\rm L}. 
\label{15}
\end{equation}

Here, we assume $T_{\rm R}=\frac{\alpha+1}{\alpha-1}T_{\rm L}+\delta$. Then, Eq. \ref{14} is 
simplified as 
\begin{equation}
(\alpha-1)AC\delta \left(\frac{2\alpha}{\alpha-1}T_{\rm L}+\delta \right)^{\alpha-1}=2. 
\label{16}
\end{equation}
When $X\equiv\frac{2\alpha}{\alpha-1}T_{\rm L}\gg \delta$ ($T_{\rm L} \gg \frac{\alpha-1}{2\alpha}\delta$), 
the term $(\frac{2\alpha}{\alpha-1}T_{\rm L}+\delta)^{\alpha-1}$ is expanded around $\delta=0$ using 
Taylor expansion as, 
\begin{equation}
\begin{split}
(X+\delta)^{\alpha-1}=&X^{\alpha-1}+(\alpha-1)X^{\alpha-2}\delta \\
&+\frac{1}{2}(\alpha-1)(\alpha-2)X^{\alpha-3}\delta^2+\cdots. 
\label{17}
\end{split}
\end{equation}
Then, Eq. \ref{16} becomes 
\begin{equation}
(\alpha-1)AC\delta \left(\frac{2\alpha}{\alpha-1}T_{\rm L}\right)^{\alpha-1}+\mathcal{O}(\delta^2)=2. 
\label{18}
\end{equation}
Ignoring an order of $\delta^2$ [$\mathcal{O}(\delta^2)$], 
\begin{equation}
\delta \cong \frac{2}{(\alpha-1)AC} \left(\frac{2\alpha}{\alpha-1}T_{\rm L}\right)^{1-\alpha}, 
\label{19}
\end{equation}
is obtained. 
Thus, $T_{\rm c}$ becomes
\begin{equation}
T_{\rm c} \equiv T_{\rm R} (k=0) \cong \frac{\alpha+1}{\alpha-1}T_{\rm L}+\frac{2}{(\alpha-1)AC} \left(\frac{2\alpha}{\alpha-1}T_{\rm L}\right)^{1-\alpha}. 
\label{20}
\end{equation}
The condition $T_{\rm L} \gg \frac{\alpha-1}{2\alpha}\delta$ becomes 
\begin{equation}
T_{\rm L} \gg \left(\frac{2}{(\alpha-1)AC}\right)^{\frac{1}{\alpha}} \left(\frac{\alpha-1}{2\alpha}\right), 
\label{21}
\end{equation}
by using Eq. \ref{19}. 
Thus, Eq. \ref{20} becomes 
\begin{equation}
\begin{split}
\frac{T_{\rm c}}{T_{\rm L}} \cong \frac{\alpha+1}{\alpha-1}+\frac{2}{(\alpha-1)AC} \left(\frac{2\alpha}
{\alpha-1}T_{\rm L}\right)^{-\alpha}\to \frac{\alpha+1}{\alpha-1}, 
\label{22}
\end{split}
\end{equation}
when $T_{\rm L} \gg \left(\frac{2}{(\alpha-1)AC}\right)^{\frac{1}{\alpha}} \left(\frac{\alpha-1}{2\alpha}\right)$. 
This result shows that low  $T_{\rm L}$ and large $\alpha$ are necessary to realize low $T_{\rm c}$. 
Eq. \ref{15} becomes 
\begin{equation}
T_{\rm c} \cong 3T_{\rm L}+\frac{1}{2ACT_{\rm L}}, 
\label{23}
\end{equation}
when $T_{\rm L} \gg \sqrt{\frac{1}{2AC}}$, which is equal to Eq. \ref{20} at $\alpha=2$. 

Until now we saw analytical expressions of NDTR properties in the condition of Eq. \ref{21}.  
Next we would like to see analytical expressions of NDTR properties in a condition 
of $T_{\rm L} \to 0$ limit, although $T_{\rm L}=0$ is unrealistic situation. 
Thus, we can have mathematically more simple analytical expressions and scaling behaviours, 
when $T_{\rm L}=0$ is substituted in Eqs. 10-13 as, 
\begin{equation}
T_{\rm mR0}=\frac{ACT_{\rm R}^{\alpha+1}+T_{\rm R}}{2+ACT_{\rm R}^{\alpha}},
\label{24}
\end{equation}
\begin{equation}
J_0=A\left(T_{\rm R}-T_{\rm mR0}\right)=\frac{AT_{\rm R}}{2+ACT_{\rm R}^{\alpha}},
\label{25}
\end{equation}
\begin{equation}
\Delta T_0\equiv T_{\rm mR0}-T_{\rm mL0}=\frac{ACT_{\rm R}^{\alpha+1}}{2+ACT_{\rm R}^{\alpha}},
\label{26}
\end{equation}
\begin{equation}
k_0 (=r_0^{-1})\equiv \frac{\partial J_0}{\partial T_{\rm R}}=\frac{A\left(1-\frac{\alpha}{1+\frac{2}{ACT_{\rm R}^{\alpha}}}\right)}{2+ACT_{\rm R}^{\alpha}}, 
\label{27}
\end{equation}
where $T_{\rm mL0}=T_{\rm R}-T_{\rm mR0}$. 
In addition, a temperature which exhibits the NDTR effect ($T_{\rm c0}$) is analytically solved as,
\begin{equation}
T_{\rm c0}=\left(\frac{2}{AC(\alpha-1)}\right)^{\frac{1}{\alpha}},
\label{28}
\end{equation}
when $T_{\rm L} \to 0$.

Hu {\it et al.} previously have derived analytical expressions of NDTR with a different way from 
our method \cite{hu2}. However, the expression is a formal solution, and is not specific. 
Compared with their work, our expressions are more specific, and easily compared with experiments. 

\begin{figure}[t]
\begin{center}
\includegraphics[width=70mm,clip]{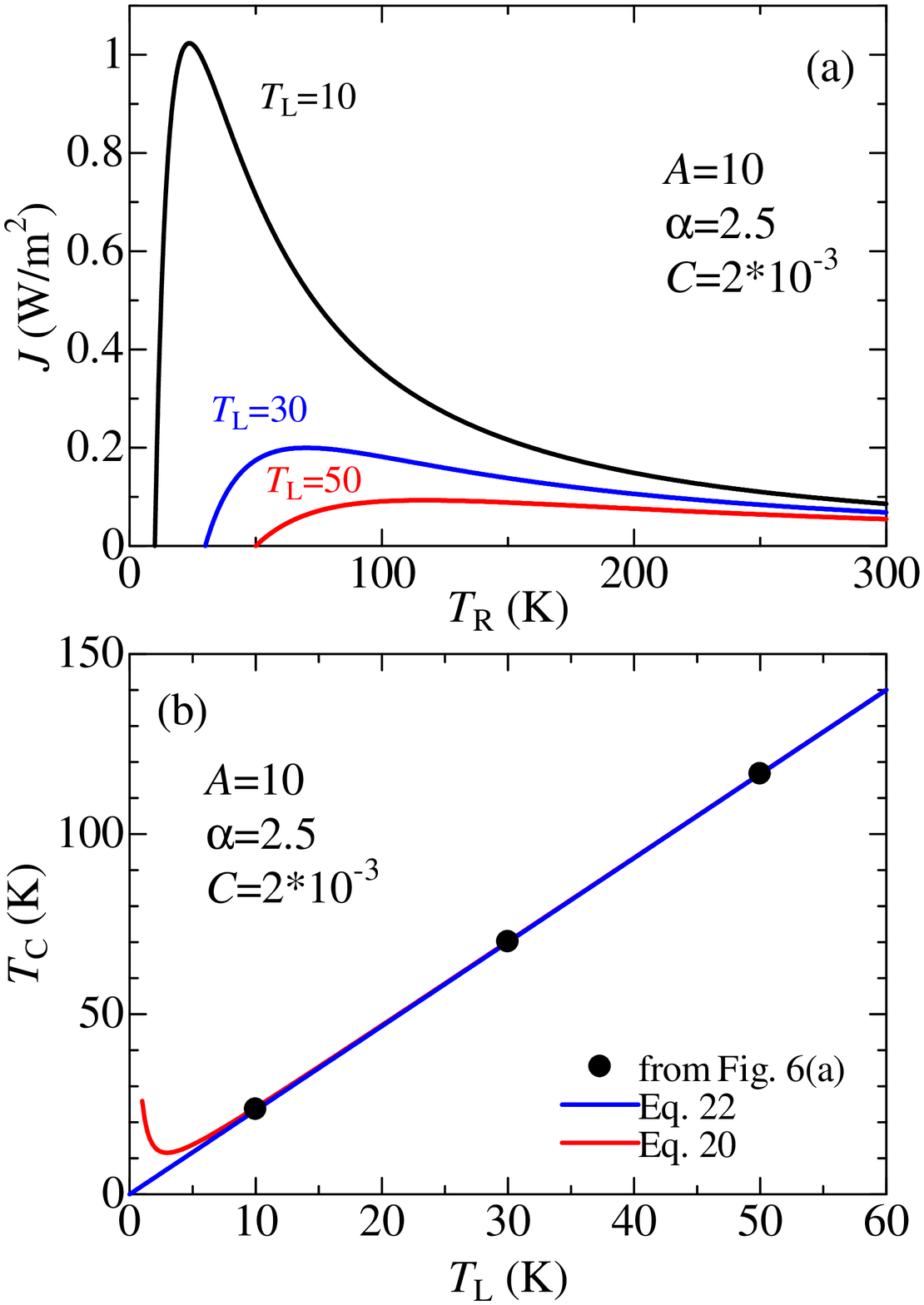}
\caption{(Color online) 
(a) $T_{\rm L}$ dependence of $J$ against $T_{\rm R}$ at $A=10$, $\alpha=2.5$, and 
$C=2\times 10^{-3}$, and (b) $T_{\rm c}$ against $T_{\rm L}$ described by 
analytical expressions Eqs. \ref{20}, 
and \ref{22} at $A=10$, $\alpha=2.5$, and $C=2\times 10^{-3}$. Dots are plotted 
from temperatures at $k=0$ of the data in Fig. \ref{f6}(a).
}
\label{f6}
\end{center}
\end{figure} 

\section{results and discussion}

\begin{figure*}[t]
\begin{center}
\includegraphics[width=130mm,clip]{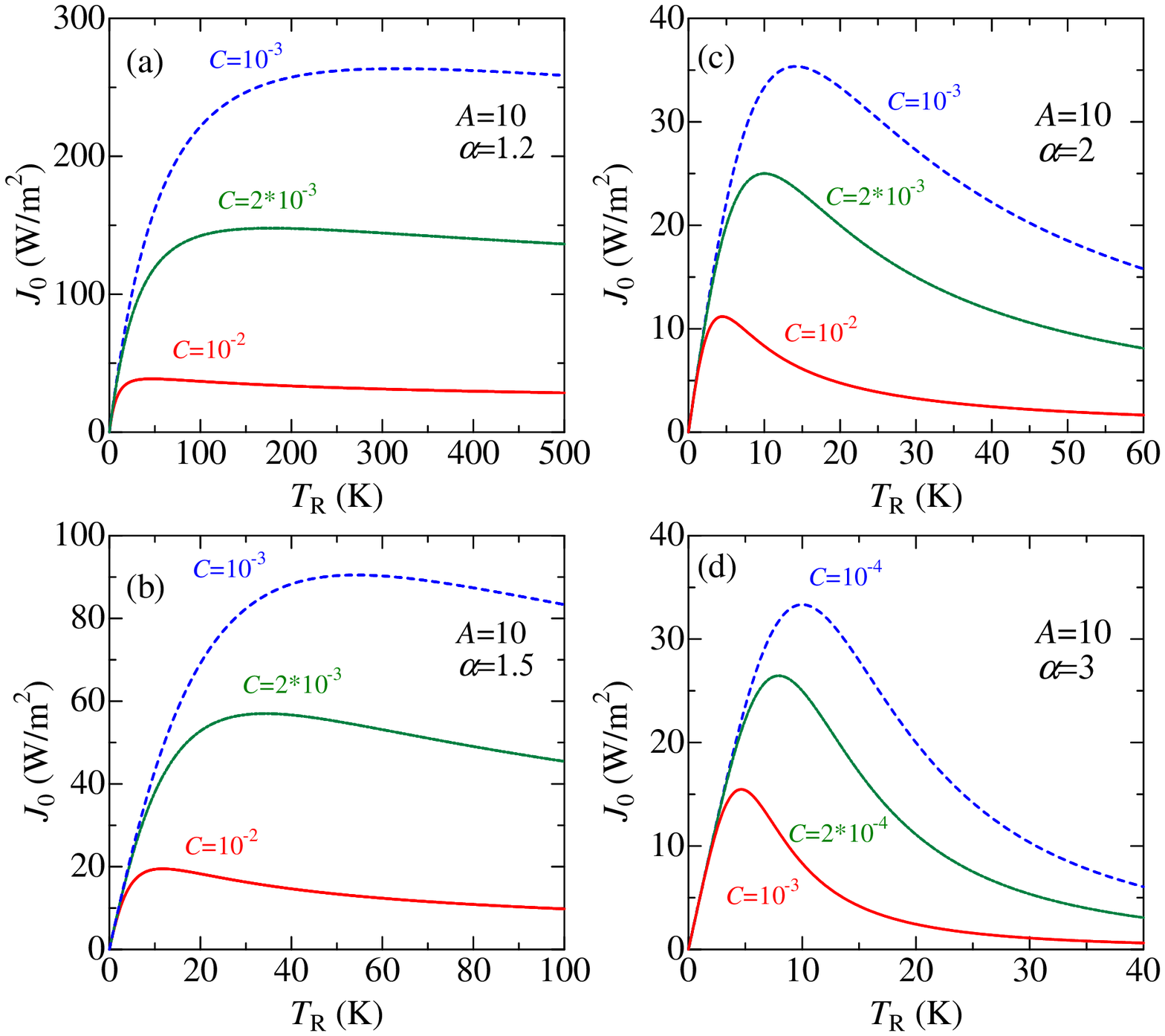}
\caption{(Color online) 
Parameter $C$ dependence of $J_0$ against $T_{\rm R}$ at (a) $A=10$ and $\alpha=1.2$, 
(b) $A=10$ and $\alpha=1.5$, (c) $A=10$ and $\alpha=2$, and (d) $A=10$ and $\alpha=3$. 
All the lines represent analytical expressions from Eq. \ref{25}. 
}
\label{f7}
\end{center}
\end{figure*} 

\begin{figure}[t]
\begin{center}
\includegraphics[width=80mm,clip]{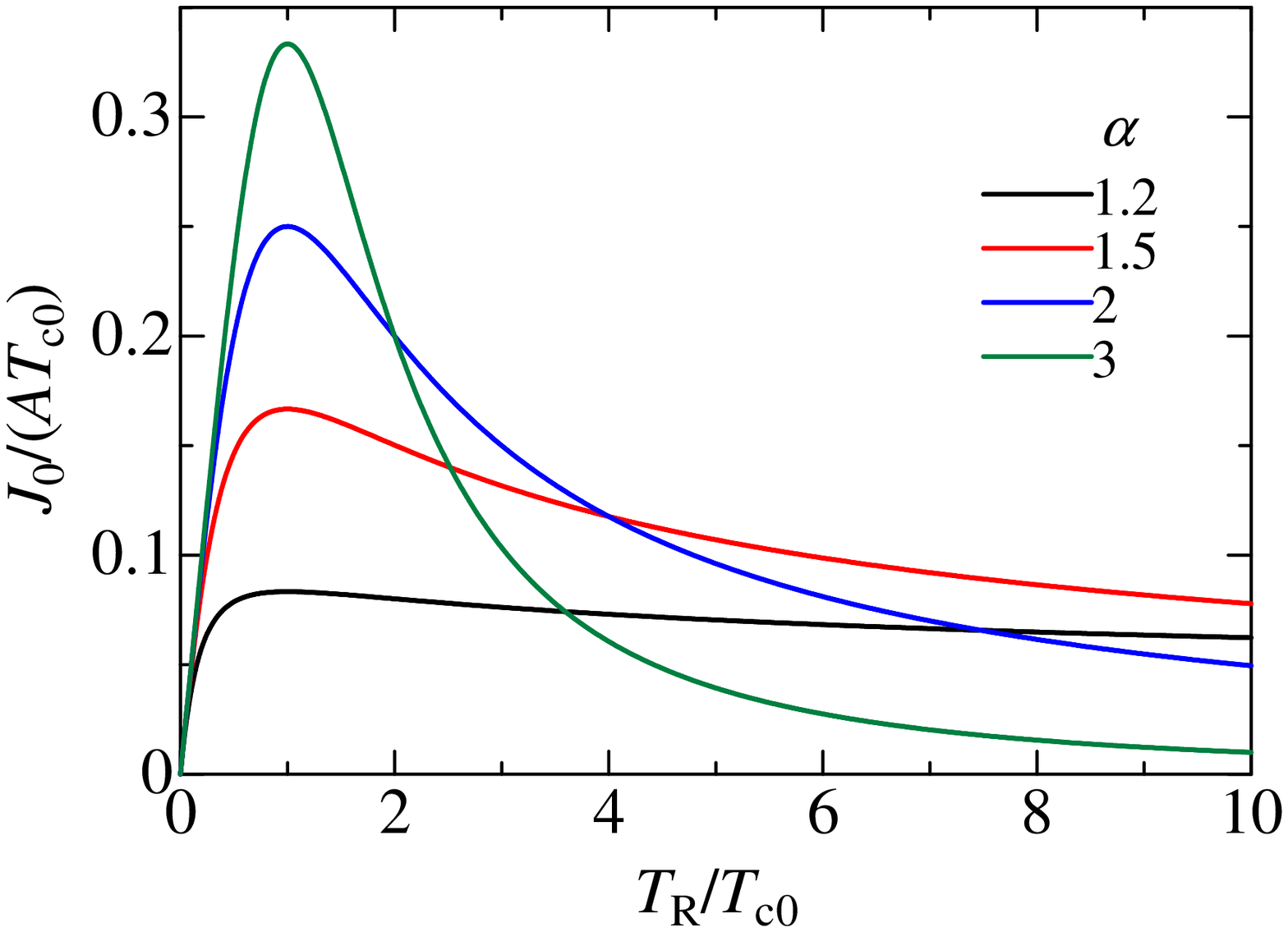}
\caption{(Color online) 
$\alpha$ dependence of dimensionless heat flux ($J'_0 \equiv \frac{J_0}{AT_{\rm c0}}$) against 
dimensionless temperature ($T' \equiv \frac{T_{\rm R}}{T_{\rm c0}}$). 
All the lines represent analytical expressions from Eq. \ref{29}. 
}
\label{f8}
\end{center}
\end{figure} 

\begin{figure}[t]
\begin{center}
\includegraphics[width=80mm,clip]{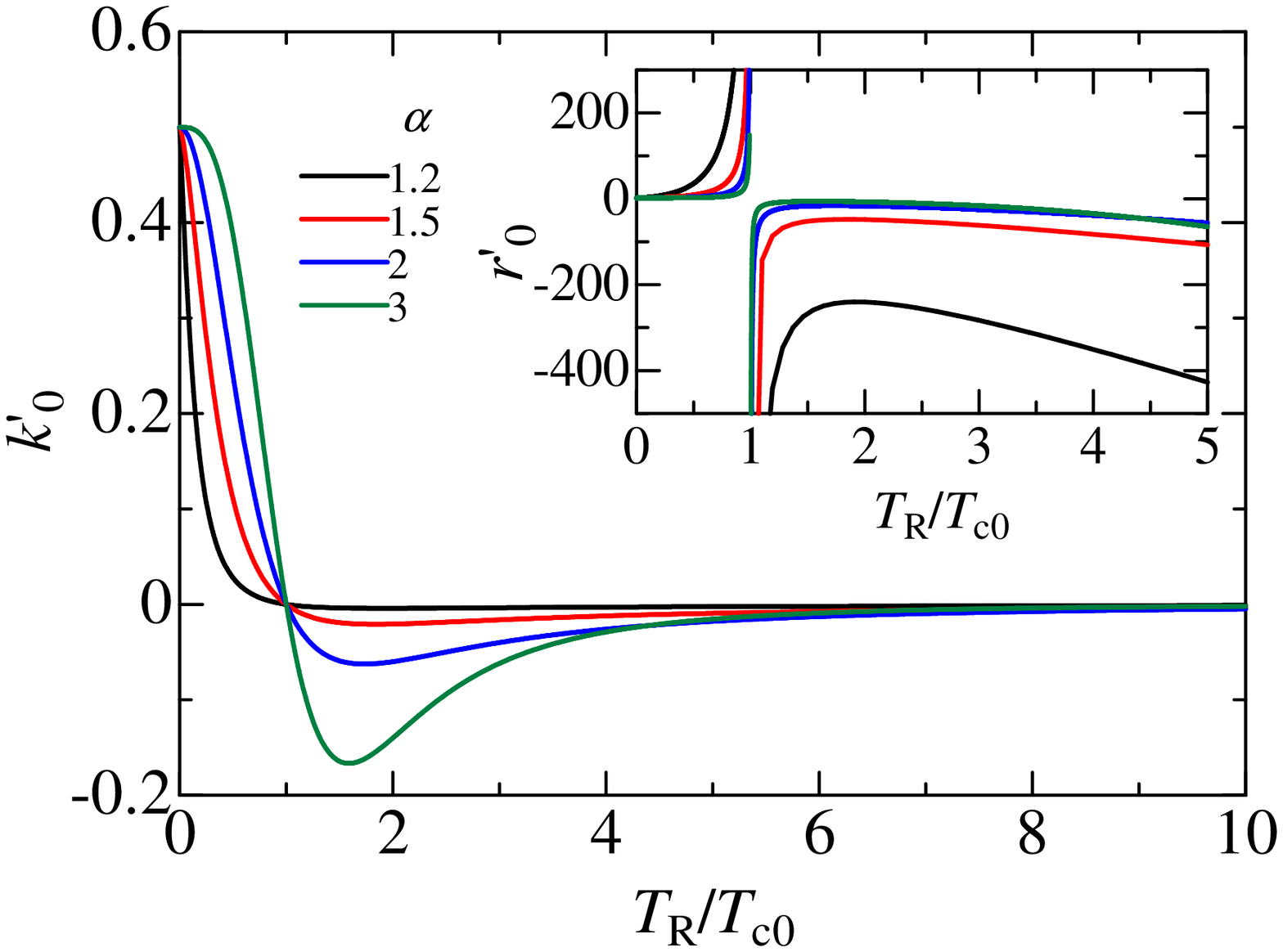}
\caption{(Color online) 
$\alpha$ dependence of dimensionless differential thermal conductance 
($k'_0 \equiv \frac{\partial J'_0}{\partial T'}$) against $T'$. 
The inset shows dimensionless differential thermal resistance ($r'_0 \equiv k'^{-1}_0$) 
against $T'$. All the lines represent analytical expressions from Eq. \ref{30}. 
}
\label{f9}
\end{center}
\end{figure} 

Figure \ref{f2}(a) shows $R_{\rm i}$ dependence of $J$ against $T_{\rm R}$. 
At $R_{\rm i}=0$, $J$ linearly increases. With non-zero 
$R_{\rm i}$, magnitude of $J$ decreases and the temperature dependence of $J$ changes. 
The reduced magnitude is caused by increased magnitude of $R_{\rm i}$. 
Figure \ref{f2}(b) shows $R_{\rm i}$ dependence of $\Delta T$ in between 
$T_{\rm Lm}$ and $T_{\rm Rm}$ at a heterojunction against $T_{\rm R}$. 
At $R_{\rm i}=0$, $\Delta T$ is zero against $T_{\rm R}$, which is caused by absence of $R_{\rm i}$. 
With non-zero $R_{\rm i}$, magnitude of $\Delta T$ increases. 
The increased magnitude is caused by increased magnitude of $R_{\rm i}$. 
All the magnitudes of $J$ increases and NDTR is not observed below $\alpha=1$, 
which is consistent with the theory. 

Figure \ref{f3}(a) shows $R_{\rm i}$ dependence of $J$ against $T_{\rm R}$. Above $\alpha=1$, 
all the data of $J$ shows reduction above $T_{\rm c}$. Thus, the NDTR effect is observed. 
Figure \ref{f3}(b) shows $R_{\rm i}$ dependence of $\Delta T$ against $T_{\rm R}$. 
Above $\alpha=1$, magnitude of $\Delta T$ also increases. 
The increased magnitude is caused by increased magnitude of $R_{\rm i}$. 
The NDTR effect is observed above $\alpha=1$, which is consistent with the theory. 

Figure \ref{f4}(a) shows $T_{\rm L}$ dependence of $J$ against $T_{\rm R}$ at 
$A=10$, $\alpha=2$, and $C=2\times 10^{-3}$. 
With $T_{\rm L}$, the magnitude of $J$ decreases, and a temperature ($T_{\rm c}$) which exhibits 
$k \equiv \frac{\partial J}{\partial T_{\rm R}}=0$ monotonically increases. 
The $T_{\rm L}$ dependent $T_{\rm c}$ is strictly understood as described in Method for $\alpha =2$. 
Figure \ref{f4}(b) shows $T_{\rm c}$ against $T_{\rm L}$. The lines are analytical 
expressions of $T_{\rm c}$ from 
Eqs. \ref{15}, \ref{22}, \ref{23}, and \ref{28} at $A=10$, $\alpha=2$, and $C=2\times 10^{-3}$. 
For $\alpha =2$, $T_{\rm L}$ dependent $T_{\rm c}$ is strictly solved as Eq. \ref{15}. 
Eq. \ref{15} is well approximated by Eq. \ref{22} above $\sim$20 K, and by Eq. \ref{23} above 
$\sim$5 K. 
At a limit of $T \to 0$, Eq. \ref{15} becomes equal to Eq. \ref{28}. 
Substituting $C=2\times 10^{-2}$ for $C=2\times 10^{-3}$, figure \ref{f4} (c) depicts 
$T_{\rm L}$ dependence of $J$ against $T_{\rm R}$. Due to enhancement of $R_{\rm i}$, 
the magnitude of $J$ decreases compared with that of $J$ in fig. \ref{f4}(a). 
As shown in Fig. \ref{f4}(d), Eq. \ref{15} is well approximated by Eq. \ref{22} above $\sim$6 K, 
and by Eq. \ref{23} above $\sim$2 K. The reductions of these temperatures compared 
with those in Fig. \ref{f4}(b) are explained by Eq. \ref{21}. 

Figure \ref{f5}(a) shows $T_{\rm L}$ dependence of $J$ against $T_{\rm R}$ at $A=10$, $\alpha=1.5$, and 
$C=2\times 10^{-3}$. With $T_{\rm L}$, $T_{\rm c}$ monotonically increases. 
Figure \ref{f5}(b) shows $T_{\rm c}$ against $T_{\rm L}$ from analytical expressions Eqs. \ref{20}, 
and \ref{22} at $A=10$, $\alpha=1.5$, and $C=2\times 10^{-3}$. Dots are plotted 
from the $T_{\rm c}$ values of the data in Fig. \ref{f5}(a). The dots are superimposed by Eq. \ref{20}. 

Figure \ref{f6}(a) shows $T_{\rm L}$ dependence of $J$ against $T_{\rm R}$ at $A=10$, $\alpha=2.5$, and 
$C=2\times 10^{-3}$. With $T_{\rm L}$, $T_{\rm c}$ monotonically increases. 
Figure \ref{f6}(b) shows $T_{\rm c}$ against $T_{\rm L}$ from analytical expressions Eqs. \ref{20}, 
and \ref{22} at $A=10$, $\alpha=2.5$, and $C=2\times 10^{-3}$. Dots are plotted 
from the $T_{\rm c}$ values of the data in Fig. \ref{f6}(a). The dots are superimposed by both 
Eqs. \ref{20} and \ref{22}. Thus, the NDTR behaviour at $T_{\rm R} \gg 0$ is well understood. 
Namely, $\alpha$ is the most important parameter to determine $T_{\rm c}$ value 
at $T_{\rm L} \gg 0$. 

Next, we examine the NDTR behaviour at $T_{\rm L} \to 0$, although $T_{\rm L}=0$ 
is unrealistic but mathematically interesting. First, we check $C$ and $\alpha$ 
dependences of $J_0$. 
Figure \ref{f7} shows parameter $C$ dependence of $J_0$ against $T_{\rm R}$ at (a) $A=10$ and 
$\alpha=1.2$, (b) $A=10$ and $\alpha=1.5$, (c) $A=10$ and $\alpha=2$, and (d) $A=10$ and $\alpha=3$. 
All the lines represent analytical expressions from Eq. \ref{25}. 
With $C$, all the magnitude of $J_0$ decreases, which is attributed to an increase of $R_{\rm i}$. 
With $\alpha$ and $C$, $T_{\rm c0}$ monotonically decreases as shown in Eq. \ref{28}. 

To further analyze these equations, we introduce dimensionless temperature 
$T' \equiv \frac{T_{\rm R}}{T_{\rm c0}}$, which shows scaling behavior. Substitution 
of $T_{\rm R}=T'T_{\rm c0}$ for Eq. \ref{25} yields 
\begin{equation}
J'_0 \equiv \frac{J_0}{AT_{\rm c0}}=\frac{T'}{2+\frac{2}{(\alpha -1)}T'^{\alpha}}.
\label{29}
\end{equation}
Left-hand-side term represents dimensionless heat flux where $AT_{\rm c0}$ represents 
heat flux in the material at $T_{\rm R}=T_{\rm c0}$ and $T_{\rm L}=0$. 
The right-hand-side term is a function of dimensionless temperature $T'$ and the 
power $\alpha$. Thus, the temperature dependence of $J'_0$ is only dependent on $\alpha$. 
In other words, the power $\alpha$ controls the temperature dependence of $J'_0$. 
The dimensionless thermal conductance $k'_0 (=r'^{-1}_0)\equiv \frac{\partial J'_0}{\partial T'}$ 
is easily derived as 
\begin{equation}
k'_0 =\frac{1-T'^{\alpha}}{2\left(1+\frac{1}{(\alpha -1)}T'^{\alpha}\right)^2}.
\label{30}
\end{equation}

Figure \ref{f8} shows $\alpha$ dependence of dimensionless heat flux ($J'_0$) against 
dimensionless temperature ($T'$). 
All the lines represent analytical expressions from Eq. \ref{29}. 
With increasing $\alpha$, peak structure at $T'=1$ becomes sharper. 
Thus, $\alpha$ is the most important parameter in the model at $T_{\rm L}=0$. 

Figure \ref{f9} shows $\alpha$ dependence of dimensionless differential thermal conductance 
($k'_0$) against $T'$. 
All the lines represent analytical expressions from Eq. \ref{30}. 
At a limit of $T_{\rm L}\to 0$, all the magnitude of 
$k$ is 0.5, which is represented by Eq. \ref{30}. Above $T'=1$, the value of $k'_0$ becomes negative, and 
it merges to minus zero at a limit of $T_{\rm R}=\infty$. 
The inset of figure \ref{f9} shows dimensionless differential thermal 
resistance ($r'_0 \equiv k'^{-1}_0$) against $T'$. 
All the lines represent analytical expressions from Eq. \ref{30}. 
Above $T'=1$, the value of $r'_0$ becomes negative, and 
it diverges to $-\infty$ at a limit of $T_{\rm R}=\infty$. 

Lastly, we would like to comment on the temperature dependence of ITR. 
As we used in this paper, $\alpha>1$ is essential to realize the NDTR effect. This means 
$R_{\rm i}(T_{\rm mL}, T_{\rm mR})$ must increase with $T$. However, experimental results show that 
$R_{\rm i}(T_{\rm mL}, T_{\rm mR})$ generally decreases with $T$ \cite{wu1}. 
As pointed out by Yang {\it et al.} \cite{yang2}, 
the NDTR effect can be realized using a material with negative thermal expansion due to 
thermal shrinkage characteristic to adjust the interface pressure. 
Indeed, Hohensee {\it et al.} have shown that $R_{\rm i}$ decreases with increasing pressure 
\cite{hohensee1}. 
Thus, if one can adjust the shrinkage properly, negative pressure effect would be obtained. 
The negative pressure effect with $T$ will enable the NDTR effect. 
There are many kinds of negative-thermal-expansion materials such as ZrW$_2$O$_8$ \cite{mary1}, 
rubber, siliceous faujasite \cite{attfield1}, siliceous zeolites \cite{lightfoot1}, 
and other inorganic materials \cite{takenaka1}. 
Proper combinations of these materials and positive-thermal-expansion materials would 
yield such a interface with positively-temperature-dependent ITR. 
We saw analytical expressions of $T_{\rm mL}$, $T_{\rm mR}$, $J$, $\Delta T$, $T_{\rm c}$, 
$k$, $r$, and now understand how to 
control the NDTR effect in a macroscopic heterojunction at zeroth order approximation. 
In the model, we found that $T_{\rm R}$ dependence of $J$ is analytically described 
using experimentally determinable parameters $A$, $C$, $\alpha$, $T_{\rm L}$, and $T_{\rm R}$. 
We could also derive an analytical expression of $T_{\rm c}$ at the condition $T_{\rm L} \gg 0$, which 
can control a temperature which exhibits NDTR behaviour. 
We also found that essentially only $\alpha$ controls temperature dependence of 
$J_0$ and $k_0$ at the limit of $T_{\rm L} \to 0$. In other words, $\alpha$ is the 
most important parameter for all $T_{\rm L}$ value. 
Sharper peak structure of $J$ ($J_0$) appears due to larger $\alpha$. Thus, using 
technology of interface control, large $\alpha$ would be developed to realize experimentally 
detectable NDTR behaviour. 
We believe that this NDTR effect in a macroscopic heterojunction with positive temperature 
dependent ITR can be realized in near future.

\section{conclusion}

In conclusion, we examine analytical expressions of NDTR effect to 
reveal a condition that enables experimental realization of the NDTR effect. 
Using $\frac{\kappa_{\rm A}}{L_{\rm A}}=\frac{\kappa_{\rm B}}{L_{\rm B}}=A$ 
approximation as zeroth order approximation, 
$T_{\rm mL}$, $T_{\rm mR}$, $J$, $\Delta T$, $T_{\rm c}$, 
$k$, and $r$ are analytically solved. All these NDTR parameters are described as a function of 
experimentally determinable parameters $A$, $C$, $\alpha$, $T_{\rm L}$, and $T_{\rm R}$. 
In particular, at a limit of $T_{\rm L}\to 0$, we found that dimensionless 
heat flux ($J'_0 \equiv \frac{J_0}{AT_{c0}}$) is only dependent on 
$\alpha$, in which larger $\alpha$ yields sharper peak structure of $J$. 
As shown in this work, $\alpha>1$ is essential to realize the NDTR effect. 
This positive temperature dependence of $R_{\rm i}$ could be possible 
when one uses a material with negative thermal expansion to adjust the interface pressure. 

\section{acknowledgment}
We would like to thank H. Kobayashi and S. Kobayashi for support.

\end{document}